\documentclass[twocolumn,pre,showpacs]{revtex4}
\usepackage[pdfpagemode=None,pdfstartview=FitH]{hyperref}
\usepackage[]{graphicx}

\begin{document}

\def\br{{\bf r}}

\title{Induced Orientational Effects in Relaxation of Polymer Melts}

\author{J. M. Deutsch and J. H. Pixley}
\affiliation{
Department of Physics, University of California, Santa Cruz, California 95064}

\date{\today}

\begin{abstract}
We study stress relaxation in bi-disperse entangled polymer solutions.
Shorter chains embedded in a majority of longer ones are known to be
oriented by coupling to them.  We analyze the mechanism for this both by
computer simulation and theoretically. We show that the results can be
understood in terms of stress fluctuations in a polymer melt and chain
screening.  Stress fluctuations are frozen on the relaxation time of the
longer chains, and these will induce strong orientational couplings in
the shorter chains.
\end{abstract}

\pacs{
}

\maketitle

Polymer molecules in concentrated solution or melts have very unusual
visco-elastic properties~\cite{Ferry} and are the subject of much 
research~\cite{DeGennesBook,RubinsteinColbyBook}. After shearing such a system
of long chains, the modulus has a long plateau that depends strongly
on chain length, yet the height of the plateau appears almost independent
of it. 

The most successful model in explaining the experimental data is the
``reptation" model of de Gennes~\cite{DeGennesReptation}. The initial
work concentrated on the case of a polymer in a background network such
as a gel which is most easily understood, but since then, reptation has had many applications
to polymer solutions. The idea is that topological effects 
restrict the movement of a polymer to a tube by virtue of {\em entanglements}
thereby suppressing transverse motion
for scales larger than the ``tube diameter". However chain ends can move
freely forming new tube. A polymer then has a curvilinear diffusion
coefficient inversely proportional to its length $L$, so that the time to 
form a new tube $\propto L^3$. This is also expected to be the relaxation
time in a polymer melt. 

To compare this to experimental data on the viscosity, we recall that
the elastic modulus $G(t)$ is related to the viscosity $\eta$ through
\begin{equation}
\label{eqn:eta:G(t)}
\eta \propto \int_0^\infty G(t) dt.
\end{equation}
Because the plateau is independent of $L$, this gives a viscosity
$\propto L^3$. The experimental data in the entangled regime
fits better with $L^{3.4}$. The explanation for this discrepancy
has yielded a number of theoretical explanations. The two main
reasons are both legitimate extensions of de Gennes original
idea. First, finite size effects~\cite{DoiFluct,RubinsteinFluct}, such as tube length fluctuations,
yield a larger slope for this quantity. Second, many body effects
that give rise to chain screening~\cite{ DeutschPRL,DeutschJPhysique,SemenovRubinstein} require modification
of the original single chain model. Researchers have been able to
fit experimental data separately using either idea and it is clear
that in fact both are real and experimentally contribute. 

That many body effects must be important can be seen by considering what
happens if the chains were all independent. Then density fluctuations
of the system would become very large and the system would be very
compressible. A real system has excluded volume interactions which leads to
a small compressibility and chain {\em screening}, that is, ideal chain
statistics on a large scale~\cite{DeGennesBook}.  The maintenance of
screening leads to an increase in free energy when the chain moves
out of a tube~\cite{DeutschPRL,DeutschJPhysique,SemenovRubinstein}. To
maintain constant density, this produces a stress field whose strength
grows linearly with the new tube. This implies that the motion will be
activated as the original tube is vacated. The stress is relieved by
other chains moving into or out-of the stressed regions. 

This mechanism
causes chains to follow the stress trails of other chains because as
a chain leaves a tube, it leaves a vacancy which is attractive to
other chains. The memory for this stress mechanism will decay with
a longer relaxation time than the time for the chain to vacate a
tube. This was proposed as an explanation for the difference 
between diffusion and stress relaxation~\cite{RubinsteinObukhov}.

The effects of interaction of chains can perhaps most easily be
seen by the fact that they induce partial alignment with each other. 
Many experiments have shown that there is strong induced orientational
coupling of smaller chains under stretching of a system containing
mainly longer ones,
for example in bi-disperse polymer melts~\cite{bidisperse} where
the ratio of the orientation of the smaller
chains to the longer host chains was found to be $\epsilon = 0.45$ which is 
quite large. It is the purpose of this
work to show that this phenomenon can at least partially be accounted for by
the same mechanism responsible for the many-body effects related
to chain screening mentioned above. Earlier work looking at
this by simulation techniques could reproduce similar orientational
responses by means of a direct simulation of these systems~\cite{baljon}.
This work should help to clarify the origin of this effect theoretically.

We start by performing computer simulations similar to those done
earlier on many body effects for long polymer chains~\cite{DeutschPRL,DeutschJPhysique}. 
Consider two chain lengths $L_1$ and $L_2$. 
We will assume reptation dynamics~\cite{WallMandel} for moving chains. 
Chains live on a cubic lattice in a $B \times B \times B$ box. 
In each step, the head or tail monomer of a chain is chosen at random,
and then an attempt is made to reattach it, in a random orientation, to the other end. It is rejected only 
if the new lattice position is occupied.  We assume a curvilinear diffusion
coefficient $\propto 1/L_i$, $i=1,2$, which is expected from reptation~\cite{DeGennesReptation}.
Translated to a Monte-Carlo simulation, we say that the probability of choosing a chain
is $\propto 1/L_i$. As discussed above, this system had been analyzed numerically in
considerable detail~\cite{DeutschJPhysique} and shows activated
motion in agreement with the theoretical predictions~\cite{DeutschJPhysique,SemenovRubinstein}
giving a relaxation time $\sim \exp((L/L^*)^{2/3})$ for very long chains, where
$L^*$ is a constant that depends on the density of the system. In a real melt
it would more directly depend on the plateau modulus of the melt. 

For the simulation to be relevant it must be in the
same universality class for statics and  dynamics as that of a real melt.
However the lattice model does not allow monomers to simultaneously 
rearrange themselves as
monomers reptate to  new sites. In  a real system, such local rearrangement
will happen as is required by the incompressibility of the system.
A consequence of this is that stress will be created by these rearrangements 
leading to an increase in free energy that turns out to 
be local~\cite{DeutschJPhysique}. This local increase in the free energy
is precisely what one simulates with reptation dynamics. Therefore
we do expect the same essential physical effects in these lattice 
simulations as well.

To understand memory effects in such systems, we prepare the simulation
in a sheared state, where the orientation of each monomer is now 
anisotropic. We do this by biasing the Monte-Carlo algorithm so that
it is more likely to make steps in the z-direction rather than the other two. 
The system is equilibrated in this state by letting it evolve for many
relaxation times. At this point, the effects of shearing are turned off
and directions of motion are chosen to be isotropic.

\begin{figure}[hbt]
\begin{center}
\includegraphics[width=\hsize]{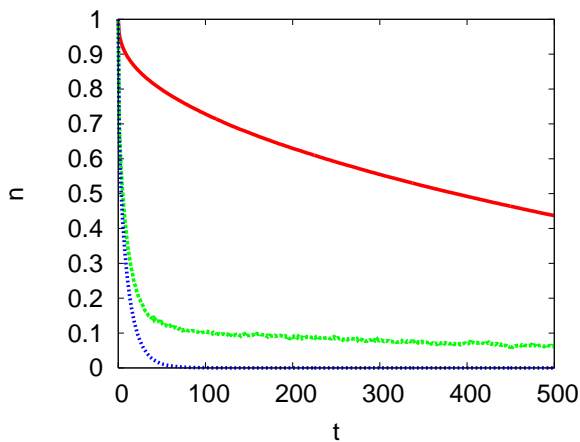}
\caption{ (Color Online) The decay of the average birefringence
for a simulation using reptational motion in a polymer melt
with two different chain lengths. The top plot, $L_2 = 256$,
and the middle plot $L_1 = 64$. The lower plot is the average
amount of $L_1$ length chain left in the tube at time $t$.
}
\label{fig:AnisotropyDecay}
\end{center}
\end{figure}

To illustrate the effect, we shall analyze what happens in a polydisperse 
system made up predominantly of long chains $L_2$ with a minority of
shorter chains $L_1$.  For example, we consider a box of size $B=16$ with
$9$ chains of length $L_2 = 256$ and one chain of length $L_1 = 64$, with a
number density of $\approx 0.58$. 
We monitor two kinds of quantities. First we monitor $l(t)$, the average
amount of the original tube left at time $t$. We also measure the birefringence
which is the deviation of polymer links from an isotropic state. We define
this as 
\begin{equation}
n \equiv 2 n_z-n_x-n_y
\end{equation}
where $n_x$, $n_y$, and $n_z$ are the
fraction of monomers going in the $x$, $y$, and $z$ directions respectively.

In the Doi Edwards model, the number of entanglement points still remaining
after time $t$ is a measure of the stress of the system. Hence in their model
the length of original tube $l(t)$ is proportional to the elastic modulus.
However if there are collective effects, even after a chain has left
its original tube, it can still have a residual orientational bias. Therefore
we should monitor more directly, the birefringence of the system, which
for small forcing will be linear will the stress of the system.

\begin{figure}[hbt]
\begin{center}
\includegraphics[width=\hsize]{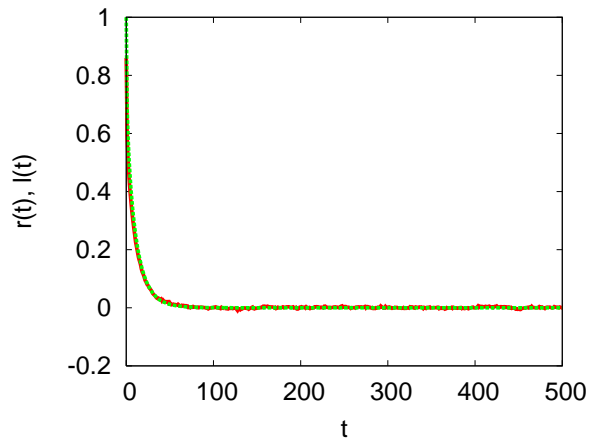}
\caption{ (Color Online) 
The residual birefringence $r(t)$ and the average
amount of chain left in a tube $l(t)$ are shown. As is
clear, the two curves closely follow each other.
}
\label{fig:bias}
\end{center}
\end{figure}

In Fig. \ref{fig:AnisotropyDecay} we plot the decay of the birefringence
separately for the longer chains (upper plot) and the shorter chain (middle plot).
The lower plot is $l(t)$, the amount of chain left in the original tube. 

In the original reptation picture, one expects that the birefringence should
be proportional to the chain left in the tube. Clearly for the shorter
chains, this is not the case. The birefringence quickly decays initially, following
$l(t)$ quite closely, however it appears to level off at a non-zero value. After
this is decays, but at a much longer time scale. We will analyze what mechanism
gives rise to such a striking effect.

In the initially biased preparation of the system, links are more likely to
be oriented in the $z$ direction, therefore as the polymer chain leaves its
tube, it will see a biased environment. In the extreme case of an environment
oriented as rods in the $z$ direction, the chain will also be biased in the
same direction as well~\cite{DeutschWarkentin}. In the case of smaller
orientational biases, it is then expected to have an effect which is
linear in the average
birefringence of the chains. If the surrounding chains are much longer
than the shorter one, it will see an almost constant orientational
bias as it moves. Even after it has completely left its original
tube it will possess a residual orientational bias. This will then
slowly decay with time at the same rate as the bias of the longer
chains.

Therefore we should modify the non-interacting reptational hypothesis
by saying that the orientational bias of the shorter chains
is the amount of chain left in the tube at time $t$ plus the bias
due to the environment. This is precisely the same assumption
made in reptation-based treatments of this effect made previously~\cite{merrill}
So we can define the ``residual birefringence'' due to the shorter chains as 
\begin{equation}
\label{eq:r}
r(t) \equiv n_1(t) - \epsilon n_2(t)
\end{equation} 
where $\epsilon$ is the fraction of orientational coupling induced by
the longer chains, as has been measured experimentally~\cite{bidisperse}. After subtracting out the background
bias from the longer chains, we are left with the bias from chains that
have yet to leave their tube. We therefore expect $r(t) \propto l(t)$.
To test if this is the
case, we find the value of $\epsilon$ that best eliminates the longer
time decay, and then plot $r(t)$ and $l(t)$ as shown
in Fig. \ref{fig:bias}. It is clear that the two curves fit each
other quite well. In this case $\epsilon = 0.14$.

Therefore we have gained a good understanding of how orientation
decays in this polydisperse polymer lattice simulation. Shorter
chains continue to have substantial anisotropy even after completely
leaving their original environments by the influence of the
surrounding longer chains, that still have not relaxed to
an unbiased state. 

This is in accord with earlier experimental
and theoretical work on these systems.
This orientational bias is an effect identical to what is
seen in more microscopic accurate simulation on effectively shorter
chains~\cite{baljon}. The degree of orientational coupling is
given by the factor $\epsilon$, which given the density of this
system, appears to be consistent with the value of coupling
found in the more microscopic approach~\cite{baljon}.

Now we turn to the relation between this lattice simulation and
real polymer melts. There are two idealizations of the simulation
that need to be addressed that stem from a lack of continuous translational 
symmetry for lattice models. The initial conditions from which the
system relaxes, and the inability of surrounding chains to simultaneously 
rearrange themselves so as to accommodate new chain.

First, in a real experiment, the elastic modulus
is defined as the stress relaxation observed after a small and
sudden deformation of the material. After such a deformation,
the system is in a nonequilibrium state, and the polymer chains
will then move to restore themselves to equilibrium statistics.
The simulation here is started in a deformed state and is then relaxed.
The deformed state has been obtained by equilibrating the chains
so that their directions are biased in the $z$ direction. This
is not identical to a real experiment because lattice chains cannot
be continuously deformed due to a lack of continuous translational
symmetry. A test of the validity of such a procedure is to ask
if we are in the small deformation, linear regime. We therefore 
test to determine if the relaxation of the system is linear in the 
initial birefringence. We calculated the integral of $n(t)$, normalized
by its initial value, for two
different initial biases, and found that indeed this was the case. For
example, we considered a melt of chains of length $64$ with a $0.58$
of the lattice sites occupied. When the bias was reduced by $38\%$,
this integral only changed by $2.3\%$.

Second, as mentioned above, there are no multi-monomer
moves with reptation dynamics, and so neighboring chains are unable 
to rearrange themselves around  new tube being formed. This
causes an increase in the local chain density of a region. In
reality, what happens instead is that the system is almost 
incompressible, so that stress and strain are created by the introduction
of new monomers into a region. These effects in equilibrium
have been analyzed in detail previously~\cite{DeutschJPhysique,SemenovRubinstein}
and give rise to the same behavior as in the simulations.

In the simulation described here, the reason that the polymer
chain acquires an anisotropy is due to the influence of the
surrounding medium. The surrounding medium is inhomogeneous and 
the density of the medium is frozen in on the time scale of a relaxation time
of the majority (longer) chains.  
The anisotropy of density fluctuations cause the polymer
to orient in the same direction. 
In a real melt, the analogous quantities to consider are {\em stress
fluctuations} as was analyzed by Rubinstein and Semenov~\cite{SemenovRubinstein}
when the system was isotropic. We will consider the effect of these
fluctuations in a stressed polymer melt. The network stress $\sigma(\br)$ will
be frozen in for timescales less than the relaxation time. When the end of
a chain enters a certain region, the network will deform by an amount
$\Delta u(\br)$. The excess free energy is then
\begin{equation}
\label{eq:F}
F \sim \int \Delta u(\br) \sigma(\br) d^3 \br .
\end{equation}
The typical energy of a fluctuation of order an entanglement length $\propto \sqrt{N_e}$,
must obey Boltzmann statistics and is therefore of order $k_B T$~\cite{SemenovRubinstein}. 
At the same time, the energy of such a fluctuation in a volume $V \propto N_e^{3/2}$
is
\begin{equation}
\label{eq:W}
W \sim \frac{V \sigma^2}{K}
\end{equation}
where $K$ is the (longitudinal) elastic modulus of the 
network and is $\propto k_B T/N_e$~\cite{DeGennesBook}.  

The increase in free energy in placing additional chain in a new
region is responsible for the exponential dependence of relaxation
time on chain length as has been previously analyzed~\cite{DeutschJPhysique,SemenovRubinstein}.
However what we are interested in here is
the {\em fluctuation} in the additional free energy 
that must be payed by placing an entanglement length of chain in a new region.
The average deformation $ u \sim N_e/V$ and
Using Eqs. \ref{eq:F} and \ref{eq:W}, this
gives~\cite{footnote:RubinsteinMistake}, $\delta F \sim T/N_e^{1/4}$.

We can therefore think about the elastic fluctuations as causing
a random potential with an R.M.S. $\propto N_e^{-1/4}$ that the
chain interacts with.  We now ask what
happens when the network is deformed. Suppose an affine deformation $e$
is applied that is of order $1$. The random potential will now
also be deformed, just as the density had been deformed in the
simulations. So we expect an anisotropy in the potential
also of order $k_B T N_e^{-1/4}$. Therefore a chain segment of
length $N_e$ will no longer explore its surroundings isotropically,
but have a bias of order $1/N_e^{1/4}$ in the z direction.
Therefore the birefringence of the chains in this new region will,
more generally be $\sim e N_e^{-1/4}$. 

To understand this more quantitatively, write the correlation
function for the random potential $\phi(\br)$ as
\begin{equation}
\label{eq:phi_phi}
\langle \phi({\bf r})\phi({\bf r}') \rangle = v({\bf r}-{\bf r}'). 
\end{equation}
For a single chain in this
environment, the annealed and quenched averages yield the same
results~\cite{catesball}. Averaging over realizations of the
random potential leads  a weak attraction of the
polymer to itself, which interacts with itself via the potential $v(\br)$. 

At first, this weak attraction would seem to be in contradiction with the
screening theorem~\cite{DeGennesBook} that says for long chains,
the statistics of individual chains are ideal. However we are considering
a potential that is static only on a time scale much less than the reptation
time. On that time scale, the curvilinear motion of the tube is small
and we can consider what happens if we stop chain ends completely from reptating.
Because of entanglements, chain motion is confined to a tube and
this means that we do not expect complete screening in this case.
Chain of length less than $N_e$ can move quite freely, but above that
entanglements reduce the number of degrees of freedom. Therefore
we expect that the screening of excluded volume is not complete, and
the screening should scale as it would for a melt of chains of length $N_e$.
Therefore the excluded volume parameter $\propto 1/N_e$~\cite{DeGennesBook}. 

On the scale of blobs of length of $N_e$, this leads to a net repulsive
potential $\sim (1/N_e) N_e^2/V \sim N_e^{-1/2}$, with a range of a tube
diameter.  This is exactly the same scale as the scale of the effective
attractive potential in Eq. \ref{eq:phi_phi}, because the height of $v$
is $\sim N_e^{-1/2}$ also with the range of a tube diameter.  Therefore
one expects that this weak attraction is canceled out by the excluded volume 
repulsion weakened by screening, in order to have ideal chains over large distances.

This argument can be turned around:
the fact that we must have complete screening for arbitrarily long chains,
and that screening will be imperfect for polymers confined to a tube,
implies that the polymer chain must move in a weak random potential
whose strength is such as to give complete cancellation of the second
virial coefficient. In other words, the assumption of reptation,
plus the requirement of screening, gives rise to the random potential
with the strength given above.

After stressing the system, the random potential will change so that
the correlation function is $v_s(\br-\br')$. The form of the
potential should now be anisotropic and approximately $v_s = v(\Lambda \br)$
where $\Lambda$ is the affine transformation of the unsheared coordinates
to the sheared state. In the direction of maximum extension, the range of
$v_s$ has now increased, and perpendicular to this, the range has
decreased. This clearly leads to an alignment of the chains in the direction
of maximum extension.
To calculate the local anisotropy quantitatively is not possible without
additional information. The exact form of fluctuation in Eq. \ref{eq:phi_phi}
depends on the correlations of entanglements in these systems. This, 
at the moment, is unknown. Related to this, is how the imperfect screening of
chains in tubes, discussed above, is modified by stretching. This is necessary
to determine chain statistics in the stretched state.

The $\epsilon$ factor in Eq. \ref{eq:r} is $\propto N_e^{-1/4}$. This is a
substantial amount of orientational bias on the scale of a tube diameter.
In a melt one would also expect that chain packing would also induce
further alignment locally due to local chain packing and indeed an $\epsilon
= 0.45$ has been reported experimentally. However in concentrated and
semidilute solutions, such local packing effects  would not be expected
to play a large role. In this case, we expect that the mechanism discussed
here would dominate.

In conclusion we have shown that induced orientational coupling of chains
in a concentrated polymer solution can be understood as a consequence
of the many body interactions of a polymer system undergoing reptation.
Reptational motion freezes in degrees of freedom causing incomplete
screening that must be balanced by a random potential causing chain
attraction. This random potential has its origin in elastic fluctuations
of the system~\cite{SemenovRubinstein}.  When the system is stretched,
the correlation in the random potential no longer screens interactions
completely and causes the chain to elongate in the direction of
stretch. This was analyzed directly by computer simulation using reptation
dynamics and demonstrated that this orientational bias is substantial.


\begin{thebibliography}{}
\bibitem{footnote:RubinsteinMistake} In reference \cite{SemenovRubinstein} this quantity
is calculated for an arbitrary length of chain. However the authors average the
strain over the entire volume, which is correct for the case of interest here, but
not in general.
\end{thebibliography}
\end{document}